\documentstyle[12pt]{article}
\def\ds{\displaystyle}
\def\beq{\begin{equation}}
\def\eeq{\end{equation}}
\def\bea{\begin{eqnarray}}
\def\eea{\end{eqnarray}}
\def\ba{\begin{array}}
\def\ea{\end{array}}
\def\del{\partial}
\def\t{\widetilde}
\def\G{\Gamma}
\def\O{\Omega}
\def\L{\Lambda}
\def\g{\gamma}
\def\e{\epsilon}
\def\sC{\not{\hbox{\kern-2.5pt {\rm C}}}}
\def\sY{\not{\hbox{\kern-1.5pt {\rm Y}}}}
\def\sV{\not{\hbox{\kern-1.5pt {\rm V}}}}
\def\sD{\not{\hbox{\kern-1.5pt {\rm D}}}}
\def\sF{\not{\hbox{\kern-1.5pt {\rm F}}}}
\def\osV{\not{\hbox{\kern-1.5pt {$\overline{\rm V}$}}}}

\def\CM{{\cal M}}
\def\CW{{\cal W}}
\def\CF{{\cal F}}

\def\CN{{\cal N}}

\def\CY{{\cal Y}}
\def\CC{{\cal C}}
\def\CA{{\cal A}}
\def\CF{{\cal F}}
\def\a{\alpha}

\def\i{\iota}

\def\s{\sigma}

\def\L{\Lambda}
\def\ra{\rightarrow}
\def\lra{\longrightarrow}
\def\Tr{{\rm Tr}}

\font\litfont = cmr6
\def\half{{\litfont {1 \over 2}}}
\def\IR{\relax{\rm I\kern-.18em R}}
\makeatletter
\@addtoreset{equation}{section}
\makeatother

\begin{document}
\setcounter{page}{0}
\pagestyle{empty}
\begin{flushright}
hep-th/0008149\\
CPHT-S082.0800\\
\end{flushright}
\vspace*{.5cm}
\begin{center}
\vspace{.5cm}
{\Large\bf D-brane Couplings, RR Fields}\\
\vspace{.5cm}
{\Large\bf and Clifford Multiplication }\\
\vspace*{1cm}
{\bf S. F. Hassan}\footnote{e-mail: {\tt fawad@cpht.polytechnique.fr}}
and 
{\bf Ruben Minasian}\footnote{e-mail: {\tt ruben@cpht.polytechnique.fr}}\\
\vspace{.2cm}
{\it Centre de Physique Th\'{e}orique, Ecole Polytechnique
\footnote{Unit\'{e} mixte du CNRS et de l'EP, UMR 7644}
\\
91128 Palaiseau, France}\\ 
\vspace{1cm}
{\bf Abstract}
\begin{quote}
Non-trivial configurations of Yang-Mills fields and gravitational
backgrounds induce charges on Dp-branes that couple them to lower and
higher RR potentials. We show that these couplings can be described in
a systematic and coordinate independent way by using Clifford
multiplication. In the minimal formulation, D-brane charges and RR
potentials combine into bispinors of an $SO(1,9)$ which is defined
with a flat metric and does not coincide with the space-time Lorentz
group. In a non-minimal formulation, the RR potentials combine into 
$SO(10,10)$ spinors while the space of charges is formally enlarged 
to construct $SO(10,10)$ bispinors. The formalism suggests that the 
general form of the gravitational contribution to the D-brane charges is
not modified, though the replacement of wedge product by Clifford
multiplication gives rise to new couplings, consistent with T-duality.
\end{quote}
\end{center}
\vfill
\begin{flushleft}
{Aug 2000}\\
\vspace{.5cm}
\end{flushleft}
\newpage
\setcounter{footnote}{0}
\pagestyle{plain}
\section{Introduction}

Couplings on D-branes to RR potentials played an important role in the
developments of the last few years. Since the early days of D-branes,
it has been known that by virtue of non-trivial gauge field
configurations on the brane worldvolume, a Dp-brane couples not only
to a $p+1$-form Ramond-Ramond potential, but also to lower
Ramond-Ramond potentials enabling it to carry charges corresponding to
smaller branes \cite{douglas}. It was later found that non-trivial
gravitational backgrounds can also induce charges on a Dp-brane
worldvolume coupling it to lower Ramond-Ramond potentials
\cite{bsv,GHM}. Finally, it was shown in \cite{CY, MM} that
topologically non-trivial normal bundles also induce RR charges. The
existence of these couplings can be checked both by anomaly
cancellation arguments, as well as by microscopic string computations.

We start with a very brief review of D-brane couplings. Let $f: \CW
\hookrightarrow X$ be the embedding of the D$p$-brane worldvolume into
space-time (See footnote $4$ for the notation). We will denote the
tangent bundles by ${\cal TW}$ and ${\cal T}X$, respectively. The
normal bundle $\CN(\CW,X)$ is then defined by
\beq
0 \ra {\cal TW} \  {\buildrel {f_{\ast}\!\!\! }
 \over \longrightarrow} {\cal T}X \ra \CN(\CW,X) \ra 0\,.
\label{nrmlbdle}
\eeq
The (anomalous) coupling on the worldvolume of $N$ coincident D-branes
may be written in the form
\beq
I_{WZ}^{(p)}=\int_{\CW} \CC \wedge \CY(\CM, g)\,, 
\label{inflow}
\eeq
where $\CW$ is the $p+1$-dimensional worldvolume of the brane,
$\CC=f^{\ast}C$ is the pullback of the total RR potential $C$, ${\cal
M} = \CF - f^*B$ where $B$ is the NS-NS 2-form potential, $\CF$ is the
field strength of the $U(N)$ gauge field on the brane, and $g$ is the
restriction of the space-time metric to the brane. Defining ${\rm ch}(
E) = {\rm tr}_N\exp\left( {\CM \over 2\pi}\right)$, the final formula
for $\CY$ is of the form 
\beq
\CY (\CM, g) = {\rm ch}( E)  e^{\half d} \sqrt{{\widehat A}({\cal TW})
\over {\widehat A}({\cal N})}\,.
\label{charge}
\eeq
Here $d$ is a degree two class defining a $Spin^c$ structure on $\CW$
and in case of (almost) complex manifolds $d=-c_1(\CN)$.

One of the important features of (\ref{inflow}) is that it can be
written in terms of the bulk data ({\it via} the so-called $K$-theoretic
Gysin map $f_{!}$) and be brought to a form
\beq
\int_X C \wedge Y = \int_X C \wedge
{\rm ch} \left( f_{!}E  \right) \sqrt{{\widehat
A}({\cal T}X)}\,,  
\label{stinf}
\eeq
thus leading to an interpretation of the RR charges ${\rm ch} \left(
f_{!}E \right) \sqrt{{\widehat A}({\cal T}X)}$ as elements ($x =
f_{!}E$) of $K$-theory of space-time $K(X)$ \cite{MM, wik}. Moreover
as turns out similar formulae and thus similar interpretations can be
possible for the RR potentials themselves \cite{MW}.

As mentioned already, one of the outcomes of the physics of WZ
couplings (starting from refs \cite{douglas, bsv} has been the
observation that the higher brane physics is influenced by the lower
branes. This logic has an obvious drawback of not being T-duality
invariant, under which lower and higher branes can be easily
interchanged. Indeed, consider a Dp-brane carrying a Dp$'$-brane
charge for $p>p'$. Since T-duality can interchange $p$ and $p'$, one
expects that a Dp-brane should also carry charges corresponding to
higher D-branes. The form of the Dp-brane coupling to higher RR forms,
when the WZ term in the worldvolume action does not contain
gravitational interactions, was obtained in \cite{TR,M} in the static
gauge. The higher brane charges are induced by the transverse
non-Abelian scalars on the D-brane worldvolume which are T-dual to the
worldvolume gauge fields. The presence of these interactions, which
are required by T-duality can also be checked by Matrix theory
calculations \cite{TR} as well as by microscopic string calculations
\cite{MG}. It is expected that analogous couplings should also be
induced by gravitational backgrounds.

A generic formula that contains both the standard gauge WZ terms
(wedge products) as well as terms involving contractions (inner
products) was given in \cite{M} in the static gauge and is of the form
\beq
I_{WZ}^{(p)}={\rm Str} \int_{\CW} \left( \exp{  \i_{\Phi}\i_{\Phi}}
\CC\right) \wedge
{\rm ch}(E)\,,
\label{rmone}
\eeq
where $\Phi^i$ ($i=p+1,\ldots, 9$) are the transverse scalars and the
usual pull-backs $\del_\mu X^i$ are replaced by the gauge covariant
derivatives $D_\mu\Phi^i$ (as we will see later, this should also include a
normal bundle connection). $\i_{\Phi}$ denotes a contraction with
$\Phi^i$ and $(\i_{\Phi})^n \CC = {1 \over n!} \Phi^{i_1} \cdots
\Phi^{i_n} C_{i_1 \cdots i_n jkl \cdots}$. This coupling serves a dual
purpose of restoring T-duality covariance and incorporating the
D-brane scalars. One can however ask questions about the inclusion of
gravitational contributions and the possibility of a systematic and
coordinate independent description of the new D-brane couplings to RR
backgrounds, on par with the case when the inner product couplings are
absent. We will concentrate on these issues in this paper. This is, in
particular, important for understanding the influence of the new
interactions on the $K$-theory interpretation of D-brane charges.

The structure of the paper is as follows: In section 2, we use
T-duality to obtain the general form of the D-brane couplings to
RR potentials and write down a constraint on the form of the
generalized brane charges. We also point out the appearance of a
normal bundle connection and describe how a flat $SO(1,9)$
metric enters in the calculations. In section 3 we show that the
couplings can be described in a systematic and coordinate independent
way by Clifford multiplication with respect to a flat metric. The
formulation is based on a flat $SO(1,9)$ Clifford algebra. We also
discuss the from of the gravitational contributions to the generalized
charges and show that, besides reproducing the known couplings, the
formalism predicts a host of new coupings not considered before, but
expected from T-duality in the presence of gravity. In section 4 we
describe an alternative formulation based on the $SO(10,10)$ Clifford
algebra which leads to the same results. In section 5, as an
application, we consider a simple example of the new couplings which
gives rise to the gravitational version of Myer's dielectric effect
\cite{M}. Section 6 contains some comments and a summary of the
results. 

\section{Dp-brane coupling to higher RR potentials from T-duality}

In this section, following \cite{TR,M}, we start with the WZ term in
the non-Abelian D9-brane action and use T-duality to write down the
generic form of Dp-brane couplings to RR potentials. One of the main
results in these references was the realization that, besides coupling
to $C^{(p+1)}$ and lower forms, Dp-branes also couple to higher form
RR potentials through charges induced on the worldvolume by
non-trivial configurations of the transverse scalars. The treatment
below is general and applies to non-trivially embedded branes and to
D-brane charges with both Yang-Mills as well as gravitational origins,
though the explicit form of the coupling to transverse gravitational
fields will not be discussed in detail. The action obtained in this
way is in the static gauge and contains terms which are no-longer in
the standard WZ form. In the next section we give a systematic,
coordinate independent formulation of the D-brane couplings in terms
of Clifford algebra valued RR-potentials and D-brane charges. This
formulation is also valid in cases where T-duality can no longer be
used to obtain the associated Dp-branes from D9-branes, that is, in
the absence of isometries in the transverse directions. An important
feature of our formalism is that the Clifford algebra in which the
brane charges take their values is always defined with respect a flat
metric, even when the D-brane is placed in curved space. Though at
first sight this may look puzzling, a closer examination shows that
the flat-space Clifford algebra arises naturally in the T-duality
transformations of RR potentials to which the D-brane charges couple.
We will briefly describe this below, before considering the couplings
required by T-duality.

Since the appearance of a flat-space Clifford algebra is central to
our formulation of the D-brane couplings, we begin by describing how
it arises in the context of T-duality transformations of the RR
potentials, using some results in \cite{fhtwo}. Denoting the RR
field strengths by $F^{(n)}$, there exist two alternative sets of RR 
potentials $C^{(n)}$ and $C\,'^{(n)}$ defined by   
\beq
F=dC-H\wedge C\,,\qquad
C\,'=C\wedge e^{-B}\,.
\label{CC'}
\eeq
Here, $H=dB$ and we use the notation $F=\sum_n F^{(n)}$, and similarly
for the potentials. The potentials $C^{(n)}$ are invariant under
$B_{MN}$ gauge transformations while $C\,'^{(n)}$ are not. From the
form of the WZ term in the D-brane worldvolume action one can see that
it is the latter set of potentials to which D-brane charges couple.
Let us now consider the non-trivial elements of the T-duality group
$O(d,d)$ which are contained in a subgroup $O(d)\times O(d)/
O(d)_{diag}$. The action of these elements on the RR potentials was
studied in \cite{fhtwo} where it was shown that the potentials
$C^{(n)}$ transform as a Lorentz spinor and their transformation
depends on the NS-NS backgrounds $G_{MN}$ and $B_{MN}$. Furthermore,
it was shown that the transformation of $C\,'^{(n)}$ is independent of
$G_{MN}$ and $B_{MN}$ and can be obtained from that of $C^{(n)}$ by
substituting a flat metric $\hat\eta_{MN}$ for $G_{MN}$ and setting
$B_{MN}$ to zero in the transformation equations\footnote{We use a
{\it hat} to distinguish the flat metric $\hat\eta$, which appears in
the definition of the duality group $O(d,d)$, from the flat space-time
metric $\eta$.}$^, $\footnote{This independence is manifest in an
alternative approach where components of $C\,'^{(n)}$ transform as
spinors of the T-duality group $O(d,d)$.}. In other words, since the
T-duality rotations of $C\,'^{(n)}$ do not depend on the NS-NS
backgrounds, they have the same form in curved space as in flat space
(defined by the flat metric $\eta_{MN}$ and vanishing $B_{MN}$ field).
This is the origin of the flat-space Clifford algebra appearing in the
next section. Though the transformation can be directly read off from
\cite{fhone,fhtwo}, we give a simple derivation below to make the
discussion self-contained. The reader not interested in the details
can skip to equation (\ref{C'-dual}).

As stated above, to obtain the transformation of $C\,'^{(n)}$ under the
action of non-trivial elements of the T-duality group, it suffices to
consider the flat space case. The non-trivial elements of the
T-duality group can be parametrized by a matrix $R\in O(d)\subset
O(1,9)$ which acts (in flat space) on the left-moving parts of
space-time coordinates and worldsheet fermions as $\del_+\t
X=R\,\del_+X$, $\t\psi_+=R\,\psi_+$; leaving the right-moving parts
unchanged. In the Ramond sector, the zero modes of ${\sqrt 2}\psi^M_+$
and ${\sqrt 2}\psi^M_-$ satisfy the Clifford algebra $\{\G^M,\G^N\}=
2\eta^{MN}$ and the transformation of the zero modes of $\psi^M_+$ can
be re-written as $R^M_{\,N}\,\psi^N_{o+} =\O^{-1}\psi^M_{o+}\Omega$,
where $\O$ is the spinor representation of the rotation matrix $R$.
This relative rotation of the two Clifford algebras originating in the
left- and right-moving sectors of the worldsheet can be absorbed by
the corresponding spin fields $S_+$ and $S_-$ (which generate the
Ramond ground states from the corresponding Neveu-Schwarz ground
states), giving rise to their T-duality transformation,
\beq 
\t S_+=\O\, S_+\,,\qquad \t S_-= S_-\,.
\eeq
Let us now restrict ourselves to the case of a single discrete
T-duality transformation, say, along the $X^1$ direction for which
$\t\psi_+^1=-\,\psi_+^1$. In this case one can easily construct the
spinor representation as $\O=a_{(i-f)}\G_{11}\G_1$. Here, $a_{(i-f)}$
denotes a sign ambiguity that could depend\footnote{We would like to
thank A. Sen for a discussion on this issue.} on the initial theory
{\it i} and the final theory {\it f}. For a T-duality from IIA to IIB,
we choose $a_{(A-B)}=+1$ and for one from IIB to IIA, we set
$a_{(B-A)}=-1$. The relative sign between these two cases is fixed by
the requirement that T-duality squares to $+1$ on the Ramond sector,
and the overall sign is fixed such that the WZ term in the worldvolume
action does not change sign under T-duality.

Let us now consider the RR vertex operator in flat space, $V_{RR}
=\bar S^s_+{\sF}_{\circ\,ss'} S^{s'}_-$, where ${\sF}_\circ$ denotes the
RR bispinor in flat space. For the sake of definiteness, we take $S_-$
to have negative chirality in both IIA and IIB theories. Invariance of
$V_{RR}$ under T-duality leads to the transformation of ${\sF}_\circ$
which, after taking the above sign conventions into account, takes the
form $\t{\sF}_\circ = \G_1\,{\sF}_\circ$. In flat space the RR
potentials $C^{(n)}$ and $C\,'^{(n)}$ both reduce to $C^{(n)}_\circ$,
given by $F_\circ = dC_\circ$, the transformation of which under the
above T-duality can be easily worked out as $\t{\sC}_\circ =
-\G_1\,{\sC}_\circ$. Here, $\sC_\circ$ stands for the bispinor
constructed out of the flat-space RR potentials. 
Using the result for a single T-duality, the transformation of
$C_\circ^{(n)}$ under $d$ T-dualities along $X^1, X^2,\cdots, X^d$ (in
that order) can be easily written as 
\beq
\t{\sC}_\circ = (-1)^{d(d+1)/2}\,\G_1\cdots\G_d\,{\sC}_\circ\,.
\label{tdsC0}
\eeq
The transformation of the components $C^{(n)}_\circ$ can be worked out
by using the expression ${\sC}_\circ =\sum_m ((-1)^m/m!)\,
C^{(m)}_{\circ M_1\cdots M_m}\,\G^{M_1\cdots M_m}$ in (\ref{tdsC0})
and observing that the product $\G_1\cdots\G_d$ implements Hodge
duality in $d$ dimensions. Now, if the NS-NS backgrounds $G_{MN}$ and
$B_{MN}$ are switched on, the transformation of $C^{(n)}$ develops a
dependence on the backgrounds, while that of $C\,'^{(n)}$ retains the
flat-space form above. Thus we can simply replace $C^{(n)}_\circ$ by
$C^{\,'(n)}$ in the component form of the transformation and obtain 
\beq
\t C^{\,'(m+d-r)}_{\,i_{r+1}\cdots i_d\mu_1\cdots\mu_m}=
\frac{(-1)^{r(r+1)/2+d(d-1)/2+rd}}{r!}\,
C^{\,'(m+r)}_{\,i_1\cdots i_r\mu_1\cdots\mu_m} 
\e^{i_1\cdots i_r}_{{\hphantom{i_1\cdots i_r}}{i_{r+1}\cdots i_d}}
\label{C'-dual}\,.
\eeq
Note that though this equation is valid in curved space, the indices
on the {\it epsilon}-tensor are raised and lowered using a flat metric
$\hat\eta$ that appears in the definition of the T-duality group
$O(d,d)\subset O(10,10)$, and not the space-time metric $G_{MN}$. In
fact, the transformation of $C\,'^{(n)}$ in curved space can still be
written in the form (\ref{tdsC0}) provided the $\G$-matrices used are
no longer associated with the space-time Lorentz group, but are
defined with respect to an auxiliary flat metric $\hat\eta$. This
applies not only to the discrete T-dualities discussed above, but to
all non-trivial elements of the T-duality group. Though equation
(\ref{C'-dual}) could also be obtained in other ways, the derivation
presented above emphasizes the existence of an associated $SO(1,9)$
Clifford algebra defined with a flat metric. This feature will be used
in the next section.

We now get back to the derivation of Dp-brane couplings to $n$-form RR
potentials using T-duality. Following the approach in \cite{TR,M}, we
start with the WZ term in the non-Abelian D9-brane action which has no
room for coupling to higher branes\footnote{In general, unless
otherwise stated, we will use calligraphic symbols $\CC,\CY,\CA,\CF,$
{\it etc.} to denote worldvolume quantities, including pull-backs,
while the straight symbols $C,Y,A,F,$ {\it etc.} are reserved for the
bulk quantities.}
\bea
I^{(9)}_{WZ} &=& T_{(9)}{\rm Str}\int_{\CW^{(10)}}
\CC\,'\wedge \CY (\CF)\,\nonumber\\
&=&
T_{(9)}{\rm Str}\sum_n\frac{1}{n!\,(10-n)!}\,\int_{\CW^{(10)}}
\CC\,'^{(n)}_{\a_1\cdots \a_n}\, \CY^{(10-n)}_{\a_{n+1}\cdots\a_{10}}\,
d\xi^{\a_1}\wedge\cdots\wedge d\xi^{\a_{10}}\,.
\label{D9}
\eea
The coordinates $\xi^\a$ parameterize the D-brane worldvolume and
$\CC\,'=\sum_n\CC\,'^{(n)}$ contains the pull-backs to the worldvolume
of RR potentials $C\,'^{(n)}$ defined in (\ref{CC'}).
$\CY=\sum\CY^{(m)}$ denotes the lower brane charges\footnote{Of
course, the change from $C$ to $C\,'$ affects (\ref{inflow}). In order
not to clutter the notation too much, we do not introduce $\CY\,'$;
hopefully this will not lead to a confusion.} induced on the D9-brane
by worldvolume gauge fields $\CA_\a$ and background gravitational
fields $G_{MN}$ and $B_{MN}$. Explicitly, in the absence of
non-trivial gravitational contributions, $\CY (\CF)=e^\CF$, where
$\CF$ is the non-Abelian gauge field strength \cite{douglas}. In the
presence of gravitational interactions, but with $H=dB=0$, the induced
charges are given by $\CY = e^\CF\wedge \sqrt{{\widehat A}({\cal
TW})}$ \cite{GHM,CY,MM}. This expression receives $H$ dependent
contributions the exact forms of which are not known. By absorbing the
usual $e^{-B}$ factor in the RR potentials, we have lost the
manifestly gauge invariant combination $\CF-{\cal B}$, but have gained
on other fronts: First, note that now there is a clear split between
the worldvolume and space-time quantities; while $C\,'^{(n)}$ are
intrinsically space-time forms pulled back to the brane worldvolume,
$\CY^{(m)}$ are forms that live on the brane worldvolume. Second, note
that the $C\,'^{(n)}$ used above are {\it pure} Ramond-Ramond forms in
the sense that under the T-duality group they do not mix with the
NS-NS fields. This is not the case with the alternative RR potentials
$C^{(n)}$. This property will play an important role in the next
section.

Let us split the $10$-dimensional space-time coordinates $X^M$ into
the $d$ coordinates $X^i$ ($i=1,\cdots d$) and the remaining $10-d$
coordinates $X^\mu$. Assuming that the fields are independent of the
coordinates $X^i$, we perform T-dualities along these directions to
obtain the general form of the $(p=9-d)$-brane action, including its
couplings to higher RR potentials. Choosing the static gauge, the
D9-brane WZ term can be written in form suitable for T-duality as
\bea
&I^{(9)}_{WZ}&=\,\,T_{(9)}\,
{\rm Str}\sum_n\sum_r\frac{(-1)^{(n-r)(d-r)}}{n!\,(10-n)!}\,\,
{}^n C_r\,\, {}^{10-n} C_{d-r} \int_{\CW^{(10)}}
\CY^{(10-n)}_{i_{r+1}\cdots i_d\mu_{n-r+1}\cdots\mu_{10-d}}\,
\nonumber\\[.3cm]
&&\quad\times \,\,
C\,'^{(n)}_{i_1\cdots i_r\mu_1\cdots\mu_{n-r}}\, 
\e^{i_1\cdots i_d}\,dX^{1}\wedge\cdots\wedge dX^{d}
\wedge dX^{\mu_1}\wedge\cdots \wedge dX^{\mu_{10-d}}\,,
\label{D9-imu}
\eea
where ${}^n C_r$ is the combinatorial factor $n!/r!(n-r)!$. T-duality
along $d$ directions converts the D9-brane worldvolume action to a
D(9-d)-brane action and replaces all fields by their duals. Using
(\ref{C'-dual}) in (\ref{D9-imu}), and integrating over the transverse
space (assumed to be compact of volume $V_{(d)}$), one can easily
obtain the generalized WZ term for the D(p=$9-d$)-brane as
\beq
I^{(p)}_{WZ}=T_{(p)}\,{\rm Str}\sum_{s,t}\frac{1}{t!\,s!\,u!}
\int_{\CW^{(p+1)}}\t C\,'^{(t+s)}_{\mu_1\cdots\mu_t i_s\cdots i_1}
\t\CY^{(s+u)\, i_1\cdots i_s}_{{\hphantom{(s+u)\, i_1\cdots i_s}}
\mu_{t+1}\cdots\mu_{p+1}} dX^{\mu_1}\wedge\cdots\wedge 
dX^{\mu_{p+1}}\,, 
\label{D9-d}
\eeq
where $T_{(p)}=T_{(9)}V_{(d)}$ and $u=p+1-t$. It should be emphasized
that the indices $i_1\cdots i_s$ on $\t\CY$ are raised using the flat
metric $\hat\eta_{MN}$, and not the space-time metric $G_{MN}$. The
$s=0$ term is the standard WZ Lagrangian which is responsible for a
Dp-brane carrying Yang-Mills and gravity induced charges corresponding
to smaller branes. The important feature of the action, however, is
the presence of $s\neq 0$ terms, noted in \cite{TR,M} for the pure
Yang-Mills case, that couple Dp-branes to higher RR potentials. Note
that the contractions run only over the directions transverse to the
brane which, in general, need not be compact.

Below we will discuss some features of the above action in more
detail. Let $\lambda^a$ denote the set of variables on which $\CY$
could depend, for example, the gauge fields, the scalars, the metric,
the NS-NS 2-form, {\it etc}. In the above we have assumed that $\t\CY$ is
the same as $\CY$ but now expressed in terms of the dual variables,
\beq
\t\CY(\t\lambda^a)=\CY(\lambda^a(\t\lambda))\,.
\label{YtY}
\eeq
This is a constraint on the form of $\CY$ imposed by the requirement
of compatibility of the worldvolume action with T-duality. Though this
constraint is automatically satisfied in the pure Yang-Mills case, it
can be used to obtain $B$-dependent corrections in the gravitational
couplings.

Let us first consider the pure Yang-Mills case $\CY_{i_1\cdots
i_s\mu_{n-s}\cdots\mu_n}=(e^\CF)_{i_1\cdots i_s\mu_{n-s}\cdots\mu_n}$.
This case has been extensively discussed in the literature
\cite{WZTD}. Here we emphasize a subtlety regarding the identification
of the transverse scalars, which is often ignored though it is crucial
for the validity of our coordinate independent formulation of the
brane couplings in the next sections. If we start with gauge fields
$\CA_i$ with constant background values, then after T-duality we can
identify the transverse scalars on the resulting D(9-d)-brane as
$\t\Phi^j=\CA_i\,\hat\eta^{ij}$. Strictly speaking, the discussion in
the literature mostly applies to this case. However, if the $\CA_i$
depend on $X^\mu$ before T-duality, then the resulting D(9-d)-brane
after duality is non-trivially embedded in space-time and the
directions spanned by the coordinates $X^i$ are not globally
transverse to it. Therefore, the transverse scalars have to be defined
locally as sections of the normal bundle. Let $a^I_M$ ($I=9-d,\cdots
9$) span a frame in the normal bundle. Then, the transverse scalars
are defined as $\t\Phi^I=\CA_M\,\hat\eta^{MN}\,a_N^I$ and the 
$\CF_{\mu i}\,\hat\eta^{ij}$ in $\CY$ gives
\beq
D_\mu\CA_i\,\hat\eta^{ij}=\nabla_\mu\Phi^I\,a^j_I\equiv 
\left(\del_\mu\t\Phi^I+\Theta^I_{\mu J}\t\Phi^J+[\t\CA_\mu,\t\Phi^I] 
\right)a^j_I\,,
\label{nb}
\eeq
where $\Theta^I_{\mu J}=a^I_N\del_\mu a^N_J$ is the connection on the
normal bundle to the brane. The appearance of this connection, which
is required by the covariant formulation of the next section, 
can also be checked by applying T-duality to the DBI part of the
action and constructing the kinetic energy terms for $\t\Phi^I$.  
Perhaps the easiest way of seeing the above is from point of view of
supersymmetry and appearance of the connection on the normal bundle in
the context of the gauged $R$-symmetry. Similar covariantizations were
used for description of normal bundles both for M5, M2 and
D-branes \cite{fhmm}.

Now, roughly speaking, the factors of $\CF_{\mu\nu}=\t\CF_{\mu\nu}$ in
$\CY$ will combine into $e^{\t\CF}$, while $\CF_{ij}$ will give rise
to $[\t\Phi^I,\t\Phi^J]$ which will contract the indices on $\t
C\,'^{(n)}$ through the $a^i_I$. The factors $\CF_{\mu
i}\,\hat\eta^{ij}$ will couple to $C\,'{(n)}$ as a generalized
``pull-back'' which corresponds to replacing the ordinary pull-back
$\del_\mu X^i$ by $\nabla_\mu\Phi^I\,a^j_I$, as in \cite{M,MG,H},
though now with the connection $\Theta$ included. Although the
appearance of the generalized pull-back in the theory may look
appealing, it seems to suffer from two drawbacks: 1) By construction
the expression is obtained in the static gauge and is not covariant,
2) The geometric meaning of the pull-back, in terms of the embedding
of the brane in space-time, is lost. The covariant formulation in the
next sections cures both these problems and yields the generalized
pull-back in the static gauge.

While equation (\ref{YtY}) can be easily verified for the Yang-Mills
case, it can be used as a constraint for the gravitational part of the
interaction, to find correction to (\ref{charge}). First, since the metric
mixes with the antisymmetric tensor field under T-duality, it is clear
that this form should be modified when $dB\neq 0$. Second, in branes
smaller than D9-brane, the action also contains couplings which involve
contractions between the RR potentials and the transverse components
of gravitational fields, which can give raise to the gravitational
version of the dielectric effect discussed in \cite{M}. We will
discuss this in an example in section 5. In principle, once one finds
the correct $B$-dependence for the gravitational interactions of the
D9-brane, one can find the form of the corresponding terms for the
lower branes by T-duality. We will not address this issue here, but hope to
return to it in a future publication.  

\section{D-brane charges as $SO(1,9)$ bispinors - Minimal formalism}

We have seen that, besides coupling to $C\,'^{(q+1)}$ forms for $q\leq
p$, a Dp-brane also couples to $C\,'^{(q+1)}$ forms for $q>p$ through
the excitations of the transverse scalars on the worldvolume, as well
as through gravitational fields transverse to the brane. The
expression for these new couplings, obtained by T-duality, is always
in the static gauge. However, since these couplings should survive
even when the Dp-brane is not related to a D9-brane by T-duality, it
is clear that these interactions should have a more general coordinate
invariant form. Furthermore, the new interactions can no longer be
expressed neatly as a wedge product between RR potentials $C\,'$ and
the generalized brane charges $Y$. It is desirable to find a
formulation which allows us to again express the couplings as some kind
of product between the RR potentials and the complete set of brane
charges. We show that both these problems can be resolved by
considering the brane charges as taking values in a Clifford algebra
and replacing the wedge product by Clifford multiplication (by which
here we simply mean the multiplication of $\G$-matrices).

The presence of both anti-symmetrized products as well as inner
products in (\ref{D9-d}) suggests the use of Clifford multiplication
to couple $C\,'$ and $Y$, however, the metric with respect to which
the algebra is defined is not the most obvious one. The Clifford
algebra that naturally appears in the theory is the one associated
with the $SO(1,9)$ space-time Lorentz group which is defined with
respect to the space-time metric $G_{MN}$: $\{\G_M,\G_N\}=2G_{MN}$. In
string theory, the RR field strengths $F^{(n)}$ and potentials
$C^{(n)}$ come as bispinors of this Clifford algebra. However, one can
easily check that this is not the correct candidate (see footnote 6
below). The Clifford algebra that appears in the {\it minimal}
formulation of the D-brane couplings is an $SO(1,9)$ Clifford algebra
defined with a {\it flat metric} $\hat\eta_{MN}$,
\beq
\{\,\hat\G_M,\hat\G_N\,\}=2\hat\eta_{MN}\,.
\label{CAflat}
\eeq
We use a {\it hat} to emphasize the fact that we are using flat space
quantities even in curved space. Though, at first, this may look
un-natural, we noted that a result of \cite{fhtwo}, described in the
previous section, shows that such an algebra arises naturally in the
T-duality transformations of $C\,'=C\wedge e^{-B}$, which is the
quantity that appears in the D-brane worldvolume action.  We will see
below that this structure survives in the general situation even when
T-duality is not applicable. In the next section we will describe
a {\it non-minimal} formulation of the D-brane couplings in terms of
an $SO(10,10)$ Clifford algebra. The {\it minimal} formalism of this
section can also be understood as a ``gauge-fixed'' version of this
{\it non-minimal} formalism.

Let us introduce flat $SO(1,9)$ bispinors $\sC\,'$ and $\sY$
corresponding to RR potentials $C\,'^{(n)}$ and a space-time version
(the meaning of which will be made more precise below) of the generalized
D-brane charges $Y^{(n)}$,
\bea
\sC\,' &=& \sum_m \frac{(-1)^m}{m!}\, C\,'^{(m)}_{M_1\cdots M_m}\,
\hat\G^{M_1\cdots M_m}\,,
\label{Cslash}
\\
\sY &=& \sum_n \frac{(-1)^n}{n!}\, Y^{(n)}_{N_1\cdots N_n}\,
\hat\G^{N_1\cdots N_n}\,.
\label{Yslash}
\eea
The space-time quantities $Y^{(n)}$ are introduced in order to obtain
a coordinate independent description of the couplings and are defined
such that their restriction to the brane worldvolume gives the
generalized D-brane charges. For example, the worldvolume gauge fields
$\CA_\a(\xi)$ and the transverse scalars $\Phi^I(\xi)$ can be combined
into a ten dimensional space-time field $A_M(X)$ such that 
\beq
\CA_\a(\xi) = A_M(X(\xi))\,\del X^M/\del\xi^\a\,,\qquad
\Phi^I(\xi) = A_M(X(\xi))\,\hat\eta^{MN}\,a_N^I\,.
\label{Aphi}
\eeq
Here, $X^M(\xi)$ give the embedding of the brane in space-time and
$\del X^M/\del\xi^\a$ and $a^M_I$ span, respectively, the tangent and
normal bundles to the worldvolume. In general, on the worldvolume,
$Y^{(n)}_{M_1\cdots M_n}$ restricts to $\CY^{(n)\,I_1\cdots
I_s}_{\hphantom{(n)I_1\cdots I_s} \a_{s+1}\cdots\a_n}$, though we will
not give an explicit construction for the gravitational contributions.
We will return to this point later in this section.
This space-time picture is consistent both with T-duality as well as
with the $K$-theory interpretation of the D-brane charges (in the
absence of the inner product terms in the action) and gives a
covariant way of defining the generalized charges. 

The product $(\sC\,'\sY)$ can now be easily evaluated using the formula for
the product of $\G$-matrices given in the appendix, which is
essentially what we refer to as Clifford multiplication. To restrict
the resulting formula to a Dp-brane worldvolume and eliminate the
$\G$-matrices in it, we should find the correct projection operator.
To this end, note that the dimension of a D-brane worldvolume
decreases (increases) under T-dualities performed along (transverse
to) the brane. This is very similar to how RR forms pick up or lose
components under the transformation, showing that the Dp-brane
worldvolume forms transform very similar to $C\,'^{(n)}$ in
(\ref{C'-dual}). From this we infer that the Dp-brane worldvolume forms,
for all $p$, could also be combined into a flat $SO(1,9)$ bispinor, 
\beq
\sV = \sum _q \sV^{(q)}=
\sum_q\frac{(-1)^{q}}{(q)!}\, T_{(q-1)}\, 
{\rm d} X^{L_1} \wedge\cdots\wedge 
{\rm d} X^{L_q}\,\hat\G_{L_1\cdots L_{q}}\,.
\label{Vslash}
\eeq
When restricted to a given Dp-brane worldvolume defined by the embedding
$X^M(\xi^\a)$, the term $\sV^{(p+1)}$ takes the form
\beq
\sV^{(p+1)}=\frac{(-1)^{p+1}}{(p+1)!}\, T_{(p)} 
\left(\frac{\del X^{L_1}}{\del\xi^{\a_1}}\cdots \frac{\del X^{L_{p+1}}}
{\del\xi^{\a_{p+1}}}\, {\rm d}\xi^{\a_1} \wedge\cdots\wedge 
{\rm d}\xi^{\a_{p+1}}\right)\,\hat\G_{L_1\cdots L_{p+1}}\,,
\label{Vpslash}
\eeq
For convenience we have inserted the couplings $T_{(p)}$ as part of
the volume bispinor. It is now easy to write the general covariant
form of a Dp-brane coupling to all RR potentials (including the higher
ones) in terms of a Clifford product,
\beq
I_{WZ}^{(p+1)}=-{\rm Str}\int_{\CW^{(p+1)}}{\rm Tr} \left(\G_{11}
{\osV}\G_{11} \sC\,'\sY\right)\,,
\label{I-Cliff}
\eeq
where ${\rm Str}$ is the symmetrized gauge trace and ${\rm Tr}$ is a
trace over the spinor index normalized to unity.
${\osV}=\G^0{\sV}^T\G^0$ is the Dirac conjugate of ${\sV}$ and, along
with the trace, converts bispinors to forms. The factors of $\G_{11}$
are inserted to get the correct relative sign in IIA and IIB theories
and are not unexpected since we are dealing with Majorana-Weyl
bispinors. The integral over $\CW^{(p+1)}$ restricts the expression to
the Dp-brane worldvolume and picks up the contribution from the term
$\sV^{(p+1)}$ of (\ref{Vslash}) alone.

The $\G$-matrix multiplication and tracing can be carried out using
the formulae in the appendix leading to the component form for the
generalized WZ action,
\bea
\ds I_{WZ}^{(p+1)}&=&{\ds T_{(p)}{\rm Str}\sum_{s,t}
\frac{1}{s!\,t!\,u!} \int_{\CW^{(p+1)}}
C\,'^{(t+s)}_{L_1\cdots L_tN_s\cdots N_1}
Y^{(s+u)N_1\cdots N_s}_{{\hphantom{(s+u)N_1\cdots N_s}}L_{t+1}\cdots
L_{p+1}}\,}
\nonumber\\[.2cm]
&&\qquad\qquad\qquad\times\,
\ds{\del_{\a_1}X^{L_1}\cdots\del_{\a_{p+1}}X^{L_{p+1}}\, {\rm d}\xi^{\a_1}
\wedge \cdots{\rm d}\xi^{\a_{p+1}}}\,,
\label{I-Comp}
\eea
where $u=p+1-t$. The indices on $Y$ are raised using the flat metric
$\hat\eta$. Since the indices $L_1,\cdots, L_{p+1}$ are pulled-back to
the brane worldvolume, antisymmetry in the indices implies that
$N_1\cdots, N_s$ are automatically restricted to the directions
transverse to the brane. Of course, for this to happen, it is crucial
that $C\,'$ and $Y$ are contracted by the flat metric
\footnote{If one uses the curved space $\G$-matrices $\G_M$ instead of
the flat-space ones $\hat\G_M$, one still gets (\ref{I-Comp}) but now
with $C\,'$ and $Y$ contracted by the space-time metric $G_{MN}$. This
is of course incorrect since the indices $N_1\cdots, N_s$ are no
longer restricted to the transverse space. This is a reflection of the
fact that the natural bispinors appearing in the theory are the ones
constructed out of $C\,'=C\wedge e^{-B}$ and $\hat\G_M$ (as seen in
the previous section) or $C$ and $\G_M$. A bispinor containing $C\,'$
and $\G_M$ does not arise in the theory.}. In the static gauge,
$X^\mu=\xi^\mu$ and $X^i=X^i(\xi)$, (\ref{I-Comp}) reduces to
(\ref{D9-d}). Note that the interactions in (\ref{I-Comp}) explicitly
preserve the geometric nature of the embedding of the brane
worldvolume in space-time through the appearance of the pull-backs
$\del X^M/\del \xi^\a$. The generalized pull-backs discussed in
section 2 appear as part of the restriction of $D_MA_N$ to the
worldvolume which, using (\ref{Aphi}), contains
\beq
\frac{\del X^M}{\del\xi^\a}\,(D_M A_N)\hat\eta^{NP}a^I_P=
\del_\a\Phi^I + \Theta^I_{\a J}\Phi^J + [ \CA_\a, \Phi^I]\,.
\label{nbc}
\eeq
The appearance of the normal bundle connection $\Theta^I_{\a J}
=a^I_M\del_\a a^M_J$ here is a reflection of the covariance of the
formalism.

To summarize, the product $(\sC\,'\sY)$ yields a series of even (odd)
forms encoding certain couplings for IIB (IIA) theory, while
multiplication by ${\sV}^{(p+1)}$ and tracing restricts these to a
Dp-brane worldvolume. Clifford multiplication can also be formulated
abstractly, independent of $\G$-matrices, as an operation on
differential forms. Given a vector $v$ we can define a contraction
$\i_v$, using again the flat metric, and thus an isomorphism $\s_v =
v\wedge - \,\, \i_v: \L^{\rm even}X \lra \L^{\rm odd}X$ corresponding
to ${v\!\!\!/}: S^+ \otimes S^- \oplus S^-\otimes S^+ \lra S^+ \otimes
S^+ \oplus S^- \otimes S^-$. This is a Clifford multiplication by a
vector $v$, and can be straightforwardly generalized to any form
$C\,'$ with the combinatorial factors simply given by those of the
$\G$-matrix multiplication (see appendix). In this notation, 
(\ref{I-Cliff}) can be written in a more conventional form as, 
\beq
L_{WZ}^{(p+1)} = \s_{C\,'}(Y) \Big\vert_{{\cal W}_{p+1}}\,.
\label{I-CM}
\eeq
Note that $\left(\s_{C\,'}(Y)\right)_{10} = (C\,' \wedge Y)_{10}$ and
as expected there are no contractions for D9-branes. Of course, by
design (\ref{I-Cliff}) or (\ref{I-CM}) correctly reduce to the
covariant form of the coupling (\ref{rmone}) in the case of flat
space, including the normal bundle contribution to the covariant
derivatives. However, up to now we have not really specified the
``generalized brane charge" $Y$ for the curved space, other than by
outlining the generic constraints imposed on it by T-duality. When the
NS-NS $B$-field is set to zero, choosing $Y$ in the form
\beq
Y(dB=0) = {\rm ch} (x) \sqrt{{\widehat A}({\cal T}X)}\,,
\label{bzero}
\eeq
ensures that at least the part $C\,' \wedge Y$ of (\ref{I-CM})
restricts to the expected coupling (\ref{inflow}) as given by
(\ref{charge}) following the calculations of \cite{MM}. Thus we
conjecture that this form of $Y$ should not change and taking into
account the (non-Abelian) dynamics of D-branes amounts to replacing
the wedge product $C\,'\wedge Y$ by Clifford product $\s_{C\,'}(Y)$.
This form also predicts a series of new couplings containing the
curvature of the space transverse to the D-brane. It is not easy to
check these predictions explicitly for arbitrary worldvolume ${\cal
W}_q$ and arbitrary embedding, since many of the couplings have not
been calculated on the worldvolume yet. However, for simple embeddings
it shows that the transverse curvature form can enter {\it via}
contractions with the transverse components of RR fields, quite
similar to those involving the scalars, namely terms of the form
\beq
\int_{\cal W}\Tr_R {\cal R}_{pq}{\cal R}_{rs}\CC\,'^{pqrs...}\cdots.
\label{gravco}
\eeq 
Note that the flat metric is used only for the contractions, and the
(transverse) space is curved. This formula is somewhat symbolic and
appropriate restrictions to the worldvolume are assumed here.
Furthermore, there are no contractions between the transverse scalars
and the transverse components of the curvatures. We will return to
this in section 5, where the existence of such interactions will be
demonstrated more explicitly in a case of trivial embedding, and
consider a simple example of D-brane polarization where non-trivial
gravitational background and thus (\ref{gravco}) play some role.

\section{D-brane charges as $SO(10,10)$ bispinors - Non minimal
formalism} 

The formalism in the previous section is suggested by the fact that
the action of non-trivial elements $O(d)\times O(d)/O(d)_{diag}$ of
the T-duality group on the RR fields can be written in terms of
$SO(1,9)$ spinors \cite{fhtwo,fhone}. As we saw, the generalized
D-brane charges are then forms in a 10 dimensional space. One could
also try to use an alternative formalism based on the full T-duality
group $O(d,d)\subset O(10,10)$. As we will see below, a formulation in 
terms of $SO(10,10)$ spinors is possible provided the space of charges
is {\it formally} enlarged to a 20 dimensional space, though the
physical content, of course, remains 10 dimensional. For this reason
we refer to this as the non-minimal formalism.

To understand the origin of the formal enlargement of the space of
charges, we consider the Yang-Mills sector in flat space. The charges
are related to the $p+1$ gauge fields $\CA_\a$ and the $9-p$
transverse scalars $\Phi^I$, which were combined into a 10 dimensional
space-time vector $A_M$ (\ref{Aphi}). In the open string picture of a
Dp-brane, the worldvolume gauge fields are associated with the $p+1$
canonical momenta along the brane, while the transverse scalars are
associated with the $9-p$ variables, $\del_\s X$, that one could refer
to as ``windings'' or ``length densities''. However, in the bulk
closed string theory the phase space is spanned by 10 canonical
momenta $P_M$ and 10 string ``length densities'' $\del_\s X^M$. Thus
one may formally promote both the worldvolume gauge fields as well as
the transverse scalars to 10 component fields $\b{A}_M$ and
$\b{$\Phi$}^M$ (which can further combined into a 20-dimensional
vector) such that the Dirichlet boundary conditions defining the
D-brane projects these onto $\CA_\a$ and $\Phi^I$ that carry all the
physical content. This, in fact, is an alternative way of defining the
D-brane charges in a coordinate independent way.

It has been known that, in toroidal compactifications, the components
of RR potentials $C\,'^{(n)}$ can be arranged into spinors of the
T-duality group \cite{HT, fot, fhtwo}. As in the previous section,
one can now go beyond T-duality and develop a formalism based on the   
Clifford algebra associated with the $SO(10,10)$ group which, in the
off-diagonal basis, is written as
\beq
\{\g^m\,,\g^n\}=2 J^{mn}\,,\qquad 
J=\left(\ba{cc} 0 & \hat\eta \\ \hat\eta & 0 \ea\right)\,, 
\eeq
where $m,n=1\cdots 20$. Let us denote the first 10 values of the index
by $\hat m$ and the remaining 10 by $\check m$. Then, 
\beq
\{\g^{\hat m}, \g^{\check m}\} = 2 \hat\eta ^{\hat m \check n}\,.
\eeq
One can see that $\g^{\hat m}/{\sqrt 2}$ and $\g_{\hat n}/{\sqrt 2}$ 
(where, $\g_{\hat n}=\eta_{\hat n\check m}\g^{\check m}$) satisfy the
Heisenberg algebra and can be used to construct the spinor
representation of the algebra (we will not distinguish between the
raising and lowering operators and their matrix representations).
Now we combine the the RR potentials into an $SO(10,10)$ Majorana-Weyl
spinors and the volume forms into the adjoint spinor, 
\bea
|{\rm C\,'}\rangle &=& \sum_n \frac{(-1)^n}{n!}\, 
C\,'^{(n)}_{\hat m_1\cdots\hat m_n}\, \g^{\hat m_1\cdots \hat m_n}\, 
|0\rangle\,,
\\
\langle {\rm V}| &=& \sum_p \frac{(-1)^{p+1}}{(p+1)!}\, 
\langle 0|\,\g_{\hat m_{p+1}\cdots \hat m_1} \, 
d X^{\hat m_1}\wedge \cdots\wedge d X^{\hat m_{p+1}}\Big|_{\CW^{(p+1)}}
T_{(p)}\,,
\label{Vad}
\eea 
where, $dX^{\hat m}|_{\CW}=\del X^{\hat m}/\del\xi^\a d\xi^\a$. Note
that the indices on $C\,'^{(n)}$ and the space-time coordinates always
run from 1 to 10 and are not affected by the enlargement of the space
of charges. Furthermore, the D-brane charges can be combined into an
$SO(10,10)$ bispinor,
\bea
{\b{\rm Y}} &=& \sum_q\frac{(-1)^q}{q!}\,
{\b{\rm Y}}_{m_1 \cdots m_q}\, \g^{m_1\cdots m_q}\,
\\
&=& \sum_q\sum_r \frac{(-1)^q}{q!}\,{}^q C_r\,
{\b{\rm Y}}_{\hat m_1\cdots \hat m_r\check m_{r+1}\cdots 
\check m_q}\,\g^{\hat m_1\cdots \hat m_r\check m_{r+1}\cdots 
\check m_q}\,.
\eea
For example, in terms of differential forms, in the pure Yang-Mills
case we have $\b{\rm Y}= e^{\b{F}}$ with 
$\b{F}_{\hat m\hat n}=D_{[\hat m}\b{A}_{\hat n]}$, 
$\b{F}_{\check m\check n}=\{\b{$\Phi$}_{\check m},
\b{$\Phi$}_{\check n}\}$, and $\b{F}_{\hat m\check n}=
D_{\hat m}\b{$\Phi$}_{\check n}$.
The generalized WZ term can now be written as
\beq
I_{WZ}^{(p)}={\rm Str}\,\int_{\CW^{(p+1)}} \langle {\rm V}|\b{\rm Y}|
{\rm C\,'}\rangle\,.
\eeq
Only the $p$-form term in (\ref{Vad}) contributes to the integral and  
on identifying $\hat m$ with the space-time coordinate label $M$, the
component form of this expression reduces to equation (\ref{I-Comp}).

\section{An application: D0-D6 system}

We present here the simplest case when the non-trivial gravitational
background influences the D-brane polarization. In \cite{M}, D0-branes
in the background of $C_3$ potential with a constant field strength
were considered and it was shown that due to (\ref{rmone}), a system
of D0's expands into a fuzzy two-sphere. Lets consider a minimal
extension of the D2-D0 system that includes gravity. Since in essence
our discussion closely follows that of \cite{M}, we will be rather
brief here.

We consider here a system of D0-branes in the background of a
seven-form potential $C_7$ with a constant field strength and a curved
four-dimensional manifold $X$, let say $X=K3$ for sake of
concreteness. We take the field strength of the $C_7$ to be of the
form
\beq
F_{0ijkpqrs} = \e_{ijk} \e_{pqrs}
\label{sevenfs}
\eeq
with $i,j,k = 1,2,3$ and the indices $p,q,r,s = 1,2,3,4$ spanning $X$.
Following the general discussion of section 3 
(see (\ref{gravco})), we can consider
a coupling of D0-branes to the RR seven-form potential
\beq
\Tr \left(\Phi^i \Phi^j \Phi^k \right) \Tr_R R^{pq} R^{rt} C_{ijkpqrt}
\label{dzc}
\eeq

Note that the embedding (the normal bundle) is trivial and we are not
distinguishing between the pull-backed and bulk quantities; we are
also dropping the primes on RR since $B$-field is turned off. These
couplings can be seen explicitly by starting from a D6-brane with a
worldvolume $\CW_7 = \CW_3 \times X$ in a presence of a RR three-form
field. Since this is the only RR filed switched on, there is just a WZ
coupling on the brane
$$
\int_{\CW} \CC_3 \wedge \Tr_R R^2
$$
that can be T-dualized along $X$ to a more familiar D2-brane
transverse to $X$ (even if, like for $X=K3,$ there are no isometries,
we can follow our rules for T-duality). Recalling that the action on
RR fields amounts to Hodge duality along $X$, and since $C_3$ is
transverse to $X$ it gets rotated into the seven-form (\ref{sevenfs}),
while the wedge product is replaced by contractions. Following
\cite{M} for the gauge part we arrive at (\ref{dzc}).

The rest of the story directly follows \cite{M}, and performing the
non-Abelian Taylor series expansion of the RR potential looking at the
minima of the potential we arrive at the condition
\beq
[[\Phi^i,\Phi^j],\Phi^j] +  \Tr_R R_{pq}R_{rs}\e^{pqrs} \e^{ijk}
[\Phi^j, \Phi^k] = 0. 
\label{mini}
\eeq
This has a solution in the form
$$
[\Phi^i, \Phi^j] = P_X \e^{ijk} \Phi^k
$$
where $P_X= \Tr_R R_{pq}R_{rs}\e^{pqrs}$. Thus as expected we get and
equation for a fuzzy two-sphere, however we see that its radius
depends on the curvature forms of the transverse manifold $X$.
Following the work \cite{M}, further polarization effects have been
discussed in literature. We believe it is of some interest to expand
these discussions to the cases involving non-trivial backgrounds and
embeddings.

\section{Conclusions and Discussion}

It is well known that worldvolume gauge field configurations and
gravitational backgrounds induce charges on a Dp-brane corresponding
to lower dimensional branes. These charges couple to the corresponding
RR potentials by wedge products. If the background gravity
contribution to the brane charges is neglected, then T-duality
arguments, as well as Matrix theory calculations and microscopic
string calculations show that Dp-branes can also carry charges
corresponding to higher dimensional branes by virtue of non-trivial
configurations of the transverse scalars on the brane worldvolume.

We consider generic D-brane charges, which may contain both Yang-Mills
and gravitational contributions, and employ the construction based on
T-duality to write down the form of their couplings to RR potentials.
Though the full gravitational contributions to the charges are not
explicitly constructed, their forms are strongly constrained by
T-duality covariance. It also shows that, similar to the transverse
scalars, background gravity can induce charges on the worldvolume that
couple a Dp-brane to higher RR potentials.

The generalized interactions obtained in this way have three main
features: 1) the couplings can no-longer be described in a systematic
way by a wedge product of D-brane charges and RR potentials, 2) the
couplings are obtained in the static gauge and do not have a
coordinate independent form, 3) the usual pull-backs to the
worldvolume are replaced by the covariant derivatives of the
transverse scalars. Though this is an appealing way of encoding some
worldvolume interactions, the generalized pull-back does not have a
geometric meaning in terms of the embedding of the brane. For
non-trivial embeddings of the brane, the covariant derivative of the
scalars also contains a normal bundle connection that is often missed
in the discussions of D-brane T-duality.

We show that the generalized D-brane couplings to RR potentials can be
formulated in a covariant and systematic way, essentially by replacing
the wedge product by Clifford multiplication. There are two
alternative formulations based on $SO(1,9)$ and $SO(10,10)$ Clifford
algebras. The minimal formulation is based on $SO(1,9)$ which is
defined with respect to a flat metric and does not coincide with the
space-time Lorentz group, except in the flat space. In this formalism,
the generalized D-brane charges, the RR potentials and the brane
volume forms are all combined into Majorana-Weyl bispinors of the flat
$SO(1,9)$ Clifford algebra. In the non-minimal $SO(10,10)$
formulation, the space of charges is formally enlarged so that they
fit into bispinors of the $SO(10,10)$ Clifford algebra while the RR
potentials and volume forms, each combine into $SO(10,10)$ spinors.
The appearance of both Clifford algebras can be understood in terms of
the T-duality group, though our results are independent of T-duality
and remain valid even in the absence of isometries. From the T-duality
group point of view, the minimal formalism can be regarded as a gauge
fixed version of the non-minimal formalism.

Both formalisms lead to the same coordinate invariant form for
the generalized WZ interactions on Dp-brane worldvolume,
\beq
L_{WZ}^{(p+1)} = \s_{C\,'}(Y)\,\Big\vert_{{\cal W}_{p+1}}\,,
\label{cone}
\eeq
where $\sigma_{C\,'}$ denotes Clifford multiplication by $C\,'$,
defined as an operation on forms, with respect to a flat metric.
$Y$ is a space-time version of the D-brane charges, the form of which
is explicitly known in the absence of gravitational contributions. 
The restriction of the Clifford product to a Dp-brane worldvolume then
reproduces the correct covariant form of the known couplings, 
including the effective generalized pull-backs with respect to the
covariant derivatives of the transverse scalars. The covariant
derivatives now contain both the gauge as well as the normal bundle
connection. 

The full complete gravitational contribution to $Y$ is not known
though, its form is constrained by T-duality. However, it reduces to
the known form \cite{MM} when the NS-NS 3-form field strength is set
to zero, 
\beq
Y(dB=0) = {\rm ch} (x) \sqrt{{\widehat A}({\cal T}X)}\,.
\label{ctwo}
\eeq 
All the novelty then simply comes down to replacing the wedge product
in (\ref{stinf}) by Clifford multiplication of $Y$ by RR form $C\,'$.
Then (\ref{cone}) correctly reproduces, at least formally, the known
gravitational couplings (\ref{inflow}) while encoding many
interactions that have not yet been tested directly. We have not
explicitly constructed these new interactions in the general case,
though some special situations have been discussed. Of course, it is
important to verify some of the new worldvolume terms by microscopic
calculations. Obtaining a T-duality invariant form of $Y$ with a
non-trivial $B$-field is another direction of further research. Our
results should also be useful for obtaining a full non-Abelian
$\kappa$-symmetric action for D-branes.

We would like to conclude this section with a comment on RR charges
which, in principle can be derived from the equations of motion, in a
way manifestly consistent with our formulation. We do not pursue
this here, however, in view of the formalism developed here, it is
clear that one arrives at a formula similar to the one in the
introduction compatible with the interpretation of RR charges as
element of $K$-theory of space-time. 

\vspace{-.2cm} 
\section*{Acknowledgments}

We would like to thank the Erwin Schr\"odinger Institute for
hospitality. This work was supported in part by EEC contract
HPRN-CT-2000-00122.

\vspace{-.2cm} 
\appendix
\section{${\bf\Gamma}$-matrix multiplication}

We use the metric signature $\{-1, +1,\ldots, +1\}$ and define the
$\G$-matrices such that, 
\beq
\{\G^a,\G^b\}=2\eta^{ab} \,,
\label{AppB1}
\eeq
In the Majorana-Weyl representation all $\G^a$ are real, with $\G^0$
antisymmetric and others symmetric. To fuse products of $\G$-matrices
into antisymmetrized ones, we use the identity (See for example, 
\cite{VP})
\beq
\G_{a_1\cdots a_i}\,\G^{b_1\cdots b_j}=
\sum_{k=|i-j|}^{i+j} \frac{i!\,j!}{s!\,t!\,u!}\,
\delta^{[b_1}_{[a_i}\cdots \delta^{b_s}_{a_{t+1}}\,
\G_{a_1\cdots a_t]}^{{\hphantom{a_1\cdots a_t]}}b_{s+1}\cdots b_j]}\,,
\label{AppB2}
\eeq
with
$$
s=\frac{1}{2}(i+j-k)\,,\quad t=\frac{1}{2}(i-j+k)\,,\quad 
u=\frac{1}{2}(-i+j+k)\,.
$$
In the summation, only those values of $k$ appear for which $s$, $t$
and $u$ are integers, {\it i.e.}, $k=|i-j|, |i-j|+2,\cdots, i+j-2, 
i+j$.
The trace of products of $\G$-matrices is given by
\beq
{\rm Tr}\left(\G^{a_1\cdots a_l}\G_{b_1\cdots b_k}\right)
= 2^5\,\delta_{kl}\, (-1)^{k(k-1)/2}\, k!\,\delta^{[a_1}_{[b_1}
\cdots \delta^{a_k]}_{b_k]}\,. 
\label{AppB3}
\eeq
All antisymmetrizations are with unit weight. This trace is not
normalized to unity.

\end{document}